\documentclass[aps,prl,twocolumn,amsmath,amssymb,nofootinbib,superscriptaddress,floatfix,reprint,longbibliography]{revtex4-1}
\usepackage[dvips]{graphicx}
\usepackage{latexsym}
\usepackage{amsmath}
\usepackage{amsfonts}
\usepackage{amssymb}
\usepackage{bm}
\usepackage{color}
\usepackage{txfonts}
\usepackage{float}
\usepackage{url}
\usepackage[colorlinks=true, urlcolor=blue, linkcolor=blue, citecolor=blue]{hyperref}
\usepackage{ulem}
\usepackage{physics}
\begin{document}
	\newcommand{\fig}[2]{\includegraphics[width=#1]{#2}}
	\newcommand{\la}{{\langle}}
	\newcommand{\ra}{{\rangle}}
	\newcommand{\dg}{{\dagger}}
	\newcommand{\upa}{{\uparrow}}
	\newcommand{\dna}{{\downarrow}}
	\newcommand{\ab}{{\alpha\beta}}
	\newcommand{\ias}{{i\alpha\sigma}}
	\newcommand{\ibs}{{i\beta\sigma}}
	\newcommand{\hH}{\hat{H}}
	\newcommand{\hn}{\hat{n}}
	\newcommand{\hc}{{\hat{\chi}}}
	\newcommand{\hU}{{\hat{U}}}
	\newcommand{\hV}{{\hat{V}}}
	\newcommand{\br}{{\bf r}}
	\newcommand{\bk}{{{\bf k}}}
	\newcommand{\bq}{{{\bf q}}}
	\newcommand{\mr}{\mathrm}
	\def\gsim{~\rlap{$>$}{\lower 1.0ex\hbox{$\sim$}}}
	\setlength{\unitlength}{1mm}
	\newcommand{{\vhf}}{$\chi^\text{v}_f$}
	\newcommand{{\vhd}}{$\chi^\text{v}_d$}
	\newcommand{{\vpd}}{$\Delta^\text{v}_d$}
	\newcommand{{\ved}}{$\epsilon^\text{v}_d$}
	\newcommand{{\vved}}{$\varepsilon^\text{v}_d$}
	\newcommand{\pprl}{Phys. Rev. Lett. \ }
	\newcommand{\pprb}{Phys. Rev. {B}}

\title {Charge 4$e$ superconductor: a wavefunction  approach}
\author{Pengfei Li}
\affiliation{Beijing National Laboratory for Condensed Matter Physics and Institute of Physics,
	Chinese Academy of Sciences, Beijing 100190, China}
\affiliation{School of Physical Sciences, University of Chinese Academy of Sciences, Beijing 100190, China}

\author{Kun Jiang}
\email{jiangkun@iphy.ac.cn}
\affiliation{Beijing National Laboratory for Condensed Matter Physics and Institute of Physics,
	Chinese Academy of Sciences, Beijing 100190, China}
\affiliation{School of Physical Sciences, University of Chinese Academy of Sciences, Beijing 100190, China}

\author{Jiangping Hu}
\email{jphu@iphy.ac.cn}
\affiliation{Beijing National Laboratory for Condensed Matter Physics and Institute of Physics,
	Chinese Academy of Sciences, Beijing 100190, China}
\affiliation{Kavli Institute of Theoretical Sciences, University of Chinese Academy of Sciences,
	Beijing, 100190, China}
\affiliation{New Cornerstone Science Laboratory, Beijing 100190, China}

\date{\today}

\maketitle

\begin{figure*}
	\begin{center}
		\fig{6.8in}{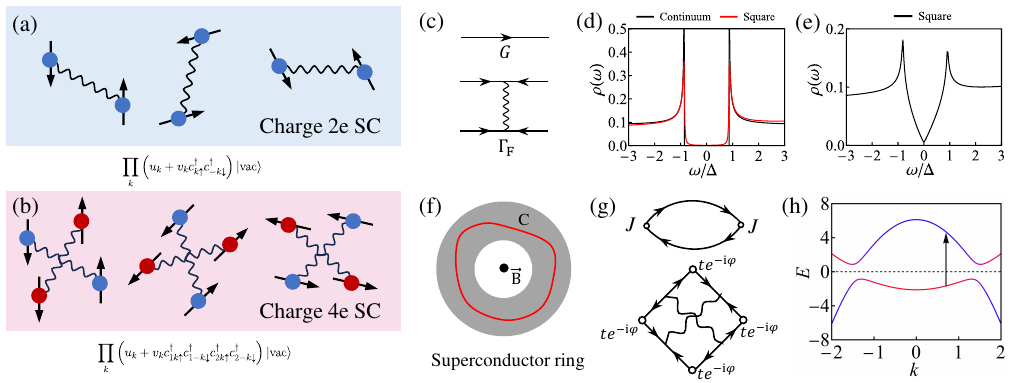}
		\caption{(a) Charge 2$e$ SC formed by two-electron Cooper pairs and its BCS wavefunction. (b) Charge 4$e$ SC formed by four-electron Cooper pairs and its wavefunction $|GS \rangle$. (c) Normal Green's function $G$ and anomalous two-particle Green's function $\Gamma_F$ of the charge 4$e$ SC. (d) U-shape DOS for charge 4$e$ s-wave pairing $\Delta_k=\Delta$ on the continuum model (black line) and the square lattice (red line).(e) V-shape DOS for charge 4$e$ d-wave pairing $\Delta_k=2\Delta(\cos k_x-\cos k_y)$ on the square lattice. The DOS is numerically calculated with normal state dispersion $\xi_k=-2t(\cos k_x+\cos k_y)-\mu$. The parameters are set as $t=1.0$,  $\mu=-3.0$, $\Delta=0.2$, $\eta=\Delta/60$. (f) Superconducting ring with a magnetic field $\vec{B}$ through the hole. A loop C (red line) is chosen for the flux quantization. (g) Top panel:The paramagnetic current-current correlation Feynman diagram. The vertex is the current operator. Bottom panel: The Josephson tunneling Feynman diagram. The vertex is the tunneling operator. (h) The optical excitation process in the charge-4$e$ SC. Unlike the charge-2$e$ case, the direct transition illustrated by the arrow is not forbidden. The red and blue line represents the electron–hole band mixing weight in the superconducting state.
			\label{fig1}}
	\end{center}
	%	\vskip-0.5cm
\end{figure*}

%\textit{Introduction}
The BCS theory of superconductivity  is one of milestones in condensed matter physics, which successfully unveils the nature of this macroscopic quantum phenomenon at a microscopic level \cite{bcs_theory,cooper}.
The essential ingredients for any superconductors (SCs) are the two-electron Cooper pairs and their phase coherence \cite{cooper}, where electrons bind together two by two and condense to form a coherent quantum state, as illustrated in Fig.\ref{fig1}(a). 
However, condensing two-electron Cooper pairs is not the only way to achieve superconductivity. Theoretically, it was proposed that the four-electron Cooper pairs can also condensate to form an SC, namely the charge 4$e$ SCs, as illustrated in Fig.\ref{fig1}(b) \cite{wucj,berg_np,Egor10,hong17}. 
Experimentally, how to realize or stabilize this charge 4$e$ superconducting state is a challenging problem. Thermal melting of pair-density-wave (PDW) order \cite{berg_np,berg_09,pdw_review}, nematic SCs \cite{jiansk} or multi-component SCs \cite{zeng2024high} is proposed to achieve this charge 4$e$ quasi-long-range order.
Other scenarios  of  realizing charge 4$e$ pairing are including interactions that favor quartetting rather than pairing  \cite{wucj} and condensing charge 4$e$ skyrmions in quadratic-band-touching systems  \cite{moon} etc.
Interestingly, a possible evidence for charge 4$e$ and even charge 6$e$ pairings are resolved recently from the Kagome superconductor CsV$_3$Sb$_5$ \cite{wangjian,kagome_review}. Using Little-Park oscillation measurement, $\frac{\Phi_0}{2}$ and $\frac{\Phi_0}{3}$ oscillations are found in CsV$_3$Sb$_5$, which was thought coming from the PDW evidence from scanning electron microscope \cite{chenhui}.

Although there is much progress in this field, how to describe this charge 4$e$ SC microscopically  and what are the electronic properties of this charge 4$e$ SC remain open questions. For instance, whether this SC is gapped and the superfluid density of it are not fully understood. 
The first attempt to solve these issues microscopically is from a quantum Monte Carlo study \cite{hong17}. Motivated by this numerical study and recent progress, we address the general microscopic properties of the  charge 4$e$ SC from a wavefunction approach in this work.

The BCS wavefunction faithfully describes the essential physical properties of SCs, especially Cooper pairs and their phase coherence. 
Inspired by it, we construct a ground state wavefunction $\ket{GS}$ for charge 4$e$ SC
\begin{eqnarray}
	|GS \rangle={\displaystyle \prod_k}(u_k+v_k e^{i\phi} c_{1k\uparrow}^\dagger c_{1-k\downarrow}^\dagger c_{2k\uparrow}^\dagger c_{2-k\downarrow}^\dagger)|0\rangle,
\end{eqnarray}
which has the similar structure as BCS wavefunction.
Here, we introduce the orbit index $1,2$ to fulfill the Pauli exclusion principle. This orbital index can be any charge 4$e$ index, like real atom orbitals, the PDW order momentum $\pm Q$, valley degree of freedom, etc.
Besides the four electron pairing term $c_{1k\uparrow}^\dagger c_{1-k\downarrow}^\dagger c_{2k\uparrow}^\dagger c_{2-k\downarrow}^\dagger$, the most important ingredient of $|GS \rangle$ is the phase  factor $e^{i\phi}$. This factor is the crucial U(1) spontaneously symmetry breaking term leading to the phase coherence of charge 4$e$ condensates as in the BCS wavefunction, which leads to the Meissner effect, Josephson effect and other SC quantum properties.
To simplify our discussion, we first choose this $\phi=0$ without losing the generality. 

Equipped with this wavefunction $|GS \rangle$, the next step is to find the mean-field Hamiltonian for this wavefunction and determine the coherence factors $u_k$, $v_k$. It is easy to prove that this wavefunction is the eigenstate of following charge 4$e$ mean-field Hamiltonian
\begin{eqnarray}
	\hat{H}_{4e}&=&\sum_{k\alpha} \xi_{\alpha k} (c_{\alpha k\uparrow}^\dagger c_{\alpha k\uparrow}+c_{\alpha -k\downarrow}^\dagger c_{\alpha -k\downarrow}) \nonumber \\
	&-&\sum_{k}(\Delta_k c_{1k\uparrow}^\dagger c_{1-k\downarrow}^\dagger c_{2k\uparrow}^\dagger c_{2-k\downarrow}^\dagger+h.c.)
\end{eqnarray}
where $\alpha=1,2$. We also choose $\xi_{\alpha k}=\xi_{k}$ for two degenerate orbitals and leave the general cases in the supplemental materials (SM).
The ground state energy is found to be $E_G=\sum_k 2\xi_{k}-E_k$, 
where $E_k=\sqrt{4\xi_k^2+\Delta_k^2}$. And the $u_k$, $v_k$ take the similar factors as the Bogoliubov–Valatin transformation treatment of BCS theory
\begin{eqnarray}
    u_k^2&=&\frac{1}{2}(1+\frac{2\xi_k}{E_k}) \\
    v_k^2&=&\frac{1}{2}(1-\frac{2\xi_k}{E_k})
\end{eqnarray}
Furthermore, the excited states can be constructed from the $|GS\rangle$ wavefunction and $\hat{H}_{4e}$ Hamiltonian. For each $k$, there are several types of excited states including single-particle excitation, pair excitation and single-hole excitation.

The Green's functions of charge 4$e$ SC can be calculated using the Lehmann representation. We start from the single particle retard Green's function $G^R(k,t)_{\alpha\beta}=-i\theta(t)\langle \{c_{\alpha k \sigma}(t),c_{\beta k\sigma}^\dagger(0)\}\rangle_{GS}$. After the Fourier transform, it is obtained as
\begin{eqnarray}
    G^R(k\sigma,\omega)_{\alpha\alpha}&=&\frac{u_k^2}{\omega+\mathrm{i}\eta-(E_k-\xi_k)}+\frac{v_k^2}{\omega+\mathrm{i}\eta+(E_k+\xi_k)}. 
    \label{green}
\end{eqnarray} 
More detailed calculations can be found in SM. Note that the two poles of the Green's function are not particle-hole symmetric counterpart, which is an important feature of charge 4$e$ SC and affects the superfluid density we will show later.
However, since the 2$e$ condensation is absence, the anomalous Green's function $F(k,t)_{\alpha\beta}=-i\langle \hat{T_t} c_{\alpha -k \sigma}^\dagger(t)c_{\beta k\sigma}^\dagger (0)\rangle_{GS}$ is always zero. The charge 4$e$ condensation  relevant anomalous Green's function is the two-particle anomalous Green's function $\Gamma_F=-\langle \hat{T_t} c_{1k \uparrow}^\dagger(t_1) c_{1 -k \downarrow}^\dagger(t_2)c_{2 k\uparrow}^\dagger(t_3)c_{2 -k \downarrow}^\dagger(t_4)\rangle_{GS}$, whose diagram is shown in Fig.\ref{fig1}(c). The Fourier transform is expressed as  $\Gamma_F(k,\omega_1,\omega_2,\omega_3)$ whose explicit form can be found in SM.

From the Green's functions, we can determine the density of states (DOS) of this system from $\rho(\omega) = -\frac{1}{\pi V}\sum_{\alpha k\sigma}\mathrm{Im}G^R(k\sigma,\omega)_{\alpha\alpha}$.
For a constant pairing potential $\Delta_k=\Delta$, an analytical expression is obtained in the SMs Eq.S6. We can find this system is a full gap system as expected. However, the gap decreases compared to charge 2$e$ SC.  At $\omega=\pm\frac{\sqrt{3}}{2}\Delta$, different from $\omega=\pm\Delta$ in the 2$e$ case, the DOS shows square root divergence. To compare with d-wave pairing with $\Delta_k=\Delta(\cos k_x-\cos k_y)$, we also carry out a numerical calculation on the square lattice. The DOS for s-wave SC $\Delta_k=\Delta$ is plotted in Fig.\ref{fig1}(d) with U-shape DOS. The DOS for d-wave SC is plotted in Fig.\ref{fig1}(e) with V-shape DOS. Hence, the charge 4$e$ is gapped or not is determined by the gap function $\Delta_k$ in our two orbital degenerate limit.
%This result is found XXX. s-wave case $\Delta_k=\Delta$, d-wave case XXX.
In the following discussion, we always choose a constant pairing $\Delta$ to capture the main feature of charge 4$e$ superconductivity.

%\textit{Flux quantization}
Regarding to the phase coherence of SC, an important prediction is the flux quantization. Taking the dynamic terms with gauge field $\mathbf{A}$ into account, the free energy for charge 4$e$ SC must contain the gauge invariant terms as
\begin{eqnarray}
	{\cal F}=\frac{\rho_s}{2} \int d\mathbf{x}  (\nabla \phi(x)-\frac{4e}{\hbar}\mathbf{A}(x))^2+F_0
\end{eqnarray}
where $\rho_s$ is the superfluid density and $F_0$ is the other irrelevant term.
The current density can be found 
\begin{eqnarray}
    \mathbf{j}(\mathbf{x})=-\frac{\delta \cal F}{\delta\mathbf{A}}=\rho_s(\frac{4e}{\hbar})^2(\frac{\hbar}{4e}\nabla\phi(x)-\mathbf{A}(x))
    \label{current}
\end{eqnarray}
In a superconducting ring with magnetic field through its hole showed in Fig.\ref{fig1}(f), we choose a loop $C$ as in Fig.\ref{fig1}(f) without any current. Then, the loop integral of Eq.\ref{current} gives $\Phi=\oint_C \mathbf{A}\cdot d\mathbf{l}=\frac{\Phi_0}{2}n$. Hence, the flux trapped in this ring is quantized as the integer multiple of flux quantum $\Phi_0=\frac{h}{4e}$, which is half of charge 2$e$ SC. This quantization condition provides a direct ``smoking gun" for charge 4$e$ SC. 

The superfluid density in Eq.\ref{current} can be obtained from the linear response theory. 
\begin{eqnarray}
    J_{\mu}=\sum_{\nu}\left[\left(-e^{2}\right) \Pi_{\mu \nu}-\frac{e^{2} \rho}{m} \delta_{\mu \nu}\right] A_{\nu}.
\end{eqnarray}
where $\rho$ is the electron density.
The $\Pi_{\mu \nu}$ is the paramagnetic current-current correlation function, which is obtained from the Kubo formula as
\begin{eqnarray}
    \Pi_{\mu \nu}(q,\omega)=-\frac{i}{V\hbar}\int_{0}^{\infty}dt e^{i\omega t} \langle [J_\mu(q,t),J_\nu(-q,0)]\rangle.
\end{eqnarray}
The calculation can be done from the Feynman diagram in the top panel of Fig.\ref{fig1}(g).
The $-\frac{e^{2} \rho}{m} \delta_{\mu \nu}$ is the diamagnetic contribution.
Using this  Kubo formula, we find the superfluid electron density $\rho_s=3\rho/4$ at zero temperature, which is different from $\rho_s=\rho$ in charge 2$e$ superconductors.
This is because the diagram contribution from the anomalous Green's function $F$ vanishes here.

The reduction of superfluid density can also be understood from the optical perspective. In charge-2$e$ SC, the optical transition between the two quasiparticle band is forbidden due to the optical selection rule from particle-hole symmetry, as proved in Ref. \cite{optical_nagaosa}. In charge-4$e$ case, however, the two poles of the Green's function Eq.\ref{green} are not particle-hole symmetric counterpart. Thus the momentum conserved transition between them exists as shown in Fig.\ref{fig1}(h), resulting in some spectral weights not being transferred to zero frequency like charge 2$e$ SC. Then the optical sum rule requires that the superfluid density must be reduced. This may be another evidence of charge 4$e$ SC for experimental verifications.

Besides the flux quantization, another key prediction of BCS theory is the Josephson effect, where Cooper pairs can tunnel between two SCs sandwiched by a tunneling barrier. We also calculate the Josephson current in our charge 4$e$ model through the Feynman diagram in the bottom panel of Fig.\ref{fig1}(g).
The DC Josephson current has the relation as $I=I_c\sin(\phi)$ where $\phi$ is the phase difference between the two SCs in the Josephson junction and the detailed calculation of critical current $I_c$ can be found in SM. Notice here, there are only four-electron Cooper pairs can tunnel through the junction since 2e condensates are zero.

More importantly, the gauge invariance also requires that $\frac{d\phi}{dt}=\frac{4e}{\hbar}V$. The AC Josephson effect with voltage bias V must be $I(t)=I_c\sin(\frac{4e}{\hbar}Vt+\phi)$.
From this expression, we can find that the AC Josephson effect measurement can be served as another hallmark for charge 4$e$ SC, where the current frequency $\omega_0$ is equal to the $\frac{4e}{\hbar}V$ \cite{hong17}.

Finally, a natural question arises: whether we can have the coexistence of charge 2$e$ and charge 4$e$ SCs. Then, the wavefunction for this case can be wrriten as
\begin{eqnarray}
	|GS \rangle_{2e-4e}={\displaystyle \prod_k}&&(u_k+v_{1k}  c_{1k\uparrow}^\dagger c_{1-k\downarrow}^\dagger+v_{2k}  c_{2k\uparrow}^\dagger c_{2-k\downarrow}^\dagger \nonumber \\
	&&+v_{4k}  c_{1k\uparrow}^\dagger c_{1-k\downarrow}^\dagger c_{2k\uparrow}^\dagger c_{2-k\downarrow}^\dagger)|0\rangle
\end{eqnarray}
This $|GS \rangle_{2e-4e}$ should be the ground state wavefunction for the charge 2$e$/4$e$ mixed hamiltonian $\hat{H}_{2e-4e}$ as
\begin{eqnarray}
	\hat{H}_{2e-4e}&=&\sum_{k\alpha} \xi_{\alpha k} (c_{\alpha k\uparrow}^\dagger c_{\alpha k\uparrow}+c_{\alpha -k\downarrow}^\dagger c_{\alpha -k\downarrow}) \nonumber\\
	&-&\sum_{k\alpha}(\Delta_k^{2e} c_{\alpha k\uparrow}^\dagger c_{\alpha-k\downarrow}^\dagger +h.c.) \nonumber\\
	&-&\sum_{k}(\Delta_k^{4e} c_{1k\uparrow}^\dagger c_{1-k\downarrow}^\dagger c_{2k\uparrow}^\dagger c_{2-k\downarrow}^\dagger+h.c.)
\end{eqnarray}
where $u_k$, $v_{1/2/4k}$ factors can be calculated numerically. The corresponding Green's functions and physical properties can be obtained following above procedures.

%\textit{discussion and summary}
In conclusion, we apply a wavefunction approach to study the charge 4e SCs. The wavefunction takes a similar structure to the BCS wavefunction for charge 2$e$ SCs. The mean-field Hamiltonian and the Green’s functions of charge 4$e$ are constructed. The physical properties including the density of states, gauge invariance, flux quantization and Josephson effects are systematically studied. All these findings provide a microscopic description of charge 4$e$ SC. We hope our work could stimulate the investigation of charge 4$e$ SC and call for further experimental tests.

\textit{Acknowledgement}
This work was supported by the National Key Basic Research and Development Program of China (2022YFA1403900), the National Natural Science Foundation of China (11888101, 12174428), the Strategic Priority Research Program of the Chinese Academy of Sciences (XDB28000000, XDB33000000), the New Cornerstone Investigator Program, and the Chinese Academy of Sciences Project for Young Scientists in Basic Research (2022YSBR-048). We thank Yi Zhou and Zixiang Li for insightful discussions.

\textit{Author contributions}
Pengfei Li, Kun Jiang, and Jiangping Hu jointly identified the problem, performed the calculations and analysis, and wrote the paper.

\end{document}